\documentclass[aps,
10pt,
prb,
twocolumn,
groupedaddress,
floatfix,
longbibliography,
superscriptaddress,
showpacs, 
showkeys,
amsfonts]{revtex4-2}

\usepackage[english]{babel}
\usepackage{amsfonts, amsmath}

\usepackage{graphicx}% Include figure files
\usepackage{dcolumn}% Align table columns on decimal point
\usepackage{bm}% bold math
\usepackage{mathrsfs} % Required for \mathscr
\usepackage{orcidlink}

\usepackage{hyperref}
\hypersetup{colorlinks=true,linktocpage=true,breaklinks=true,urlcolor=blue,linkcolor=blue,citecolor=blue}

\usepackage[
top=3cm,    % top margin
bottom=3cm, % bottom margin
left=2cm, % left margin
right=2cm % right margin
]{geometry}

\usepackage{lipsum} % Include the lipsum package

\usepackage{mathbbol}
\usepackage[sans]{dsfont}

\begin{document}
	
	\title{Quantum complexity and localization in random and time-periodic unitary circuits}
	\author{Himanshu Sahu\orcidlink{0000-0002-9522-6592}}
	\email{hsahu@perimeterinstitute.ca}
	\affiliation{Perimeter Institute for Theoretical Physics, Waterloo, ON, N2L 2Y5, Canada.}
	\affiliation{Department of Physics and Astronomy and Institute for Quantum Computing, University of Waterloo, ON N2L 3G1, Canada.}
	\affiliation{Department of Physics and Department of Instrumentation \& Applied Physics, Indian Institute of Sciences, C.V. Raman Avenue, Bangalore 560012, India.}
	\author{Aranya Bhattacharya\orcidlink{0000-0002-1882-4177}}
	\email{aranya.bhattacharya@bristol.ac.uk}
	\affiliation{School of Mathematics, University of Bristol, Fry Building, \\
		Woodland Road, Bristol BS8 1UG, UK}
	\affiliation{Institute of Physics, Jagiellonian University, Lojasiewicza 11, 30-348 Kraków, Poland. }
	\author{Pingal Pratyush Nath\orcidlink{0000-0001-5311-7729}}
	\email{pingalnath@iisc.ac.in}
	\affiliation{Centre for High Energy Physics, Indian Institute of Science, C.V. Raman Avenue, Bangalore 560012, India.}
	
	\begin{abstract}
		We study the growth and saturation of complexity in Krylov basis in random quantum circuits. In Haar-random unitary evolution, we show that, for large system sizes, this notion of complexity grows linearly before saturating at a late-time value of $d/2$, where $d$ is the Hilbert space dimension, at timescales $\sim d$. Our numerical analysis encompasses two classes of random circuits: brick-wall random unitary circuits and Floquet random circuits. In brick-wall case, complexity in Krylov basis exhibits dynamics consistent with Haar-random unitary evolution, while the inclusion of measurements significantly slows its growth down. For Floquet random circuits, we show that localized phases lead to reduced late-time saturation values of the complexity enabling us to probe the transition between thermal and many-body localized phases.
	\end{abstract}
	
	\maketitle
	
	\section{Introduction}

	Understanding the complexity of quantum states and operators is relevant to a wide range of settings, from quantum many-body physics through quantum gravity to quantum computation\,\cite{Nandy:2024htc, Rabinovici:2025otw, PhysRevLett.123.165902,harlow_quantum_2013,patrick_hayden_black_2007,shenker_stringy_2015,shenker_black_2014,yasuhiro_sekino_fast_2008,maldacena_bound_2016,hosur_chaos_2016,roberts_chaos_2017,mcclean_barren_2018,haferkamp_efficient_2023,PhysRevA.77.012307}. In many-body physics, insights into the build-up of complexity in the time evolution of an initial local observable, known as operator growth, have inspired new ways of probing the dynamics of thermalization\,\cite{PhysRevLett.123.230606,PhysRevE.50.888}. Out-of-time-order correlators (OTOCs)\,\cite{PRXQuantum.5.010201,swingle_unscrambling_2018,lewis-swan_dynamics_2019}, a quantitative tool for measuring operator growth, obeys a dynamical bound arising from unitarity and analyticity\,\cite{maldacena_bound_2016}. It is shown that in a version of quantum gravity known as anti-de-Sitter space/conformal field theory (AdS/CFT) duality, the  black holes saturate this bound\,\cite{shenker_black_2014}. Similar to black holes, OTOCs saturates the bound in models such as the so-called Sachdev-Ye-Kitaev (SYK) model which has given rise to holographic dual description with black holes\,\cite{PhysRevD.94.106002,polchinski_spectrum_2016,PhysRevLett.125.130601}.  
	
	In quantum computing, the complexity of pure states is defined as the size of the smallest circuit that produces the state from a product state, while the complexity of a unitary  is defined as the smallest circuit that approximates the unitary. This notion of quantum circuit complexity has recently gained interest due to connections between gate complexity and holography in AdS/CFT correspondence\,\cite{https://doi.org/10.1002/prop.201500092,PhysRevD.90.126007,PhysRevD.93.086006,PhysRevD.97.086015,PhysRevLett.116.191301}. It is conjectured that in the bulk theory, the wormhole's volume is dual to the boundary state's quantum complexity, whose growth has been proved for the random unitary circuits\,\cite{haferkamp_linear_2022}.
	
	Recently, the notion of state and operator complexity in Krylov basis defined using the generator of evolution operator has been extensively studied as a probe of information scrambling\,\cite{PhysRevX.9.041017,PhysRevD.106.046007,PhysRevB.106.195125,rabinoviciOperatorComplexityJourney2021b,caputaOperatorGrowth2d2021,10.21468/SciPostPhys.13.2.037,bhattacharyaOperatorGrowthKrylov2022b,bhattacharyaKrylovComplexityOpen2023d,caputaSpreadComplexityTopological2023a,rabinoviciKrylovLocalizationSuppression2022,PhysRevD.109.066010, Bhattacharya:2024szw}. Operator complexity (known as Krylov complexity) is conjectured to grow at most exponentially in non-integrable systems and can be used to extract the Lyapunov exponent, therefore, establishing a connection with OTOCs\,\cite{barbonEvolutionOperatorComplexity2019a,PhysRevX.9.041017}. On the other hand state complexity (known as spread complexity), a generalization of Krylov complexity for quantums states, is used as a probe to study quantum chaos and topological phase transitions\,\cite{PhysRevB.106.195125,caputaSpreadComplexityTopological2023a, MBLupcoming}. Furthermore, since by construction this notion of complexity measures the delocalization of a wave function in the Krylov basis with time, it also captures the localization of the wavefunction by suppression in the complexity saturation value \cite{rabinoviciKrylovLocalizationSuppression2022}.
	
	Despite a number of investigations in varying quantum systems,
	the studies of complexity in Krylov basis with discrete-time evolution are rare\,\cite{PhysRevLett.134.030401,PhysRevB.111.014309,Sahu:2024kho,PhysRevB.104.195121}. In this Letter, we study this in various classes of random quantum circuits. Quantum circuits built from local unitary gates and local measurements implemented randomly on the system with different rates are new playgrounds for quantum many-body physics and act as tractable settings to explore universal collective phenomena far from equilibrium. These models have shed light on longstanding questions about thermalization and chaos, and on the underlying universal dynamics of quantum information and entanglement\,\cite{PhysRevX.9.031009,PhysRevLett.131.220404,PhysRevB.101.104301,PhysRevB.100.134306,PhysRevLett.125.030505,PhysRevX.10.041020,PhysRevLett.125.070606,PhysRevLett.130.220404,PhysRevX.7.031016,PhysRevX.8.021014,PhysRevX.8.021013,PhysRevX.8.031058,PhysRevX.8.031057,PhysRevX.8.041019, Beetar:2025mdz}. 
	
	In random unitary circuits (RUCs) that consists of local Haar-random unitaries, any quantum state evolves towards increasingly entangled states characterized by an extensive scaling of entanglement entropy with system volume. In presence of measurements which occur repeatedly during the evolution at a fixed rate, the system undergoes a phase transition from volume-to-area-law of entanglement entropy scaling for infrequent and frequent measurement rates, respectively\,\cite{PhysRevX.7.031016,PhysRevB.100.134306}. Another class of random quantum circuits capturing important physical insights are the time-periodic ones, namely \textit{Floquet unitary circuits} (FUCs), where each time-step is repeated with the same random instances. We study two classes of FUCs introduced in ref.\,\cite{PhysRevB.98.134204} that exhibit localized phases such as Anderson localization (AL) and many-body localization (MBL).

	In this paper, we study the complexity in Krylov basis for these two aforementioned classes of quantum circuits, namely, RUC with and without randomized measurements and FUCs (where the random realization of the first timestep is repeated in all subsequent timesteps). In these models, we first study the scaling of complexity using discrete unitary setup. In the later part of the paper where we discuss the FUCs, our study uses complexity in Krylov basis as a probe to deepen the understanding of the differences between ergodicity and localization. In quantum mechanics, a system is considered ergodic if it explores its entire Hilbert space over time equally and uniformly, effectively leading to thermalization. Mathematically, it implies that the time-averaged expectation values equal ensemble averages for almost all initial states, often expressed as 
	\begin{equation}
		\lim_{T \to \infty} \frac{1}{T} \int_0^T \langle \psi(t) | O | \psi(t) \rangle dt = \langle O \rangle_\text{thermal}.
	\end{equation} 
	On the other hand, many-body localized phases maintain insulating behavior even under interaction, characterized quantitatively by the presence of localized eigenstates, which, in contrast to ergodic states, do not equitably populate the Hilbert space. Another example of such localization is Anderson localization, a situation where disorder in a system causes quantum wave functions to become localized in space and prevents them from spreading throughout the system, leading to a lack of thermalization or ergodic behavior.

	\textit{Main Result--} We find that in RUCs, the complexity undergoes a transition from linear growth at early time to sublinear and saturates at exponentially late times in system-size. We find that the growth of complexity in Krylov basis slows down and the saturation-time increases `significantly' under randomized local measurements above a threshold rate of measurement\,\footnote{While our initial numerics suggest that this threshold is near $p=0.5$ supporting previous results on complex-uncomplex transition in \cite{Suzuki_2025}, a more precise numerical analysis shows that this value is not strictly $0.5$, but system size dependent.}. In FUCs, we find the suppression of late-time saturation value of complexity in localized phases. We argue that the saturation value of complexity is proportional to $2^{\ell_d-1}$, where $\ell_d$ is the localization-length. Therefore, our notion of complexity can be used as a probe for various phase transitions in the system dynamics, namely a) measurement-induced phase transition in RUC models and b) ergodic to localized (MBL and Anderson) phases. To this end, our work therefore provides an explicit iterative circuit realization of the complexity in the Krylov basis, which probes system-size, randomized measurement rate, and the onset of localizing phases.

	% --------------------------
	\section{Background}
	% -------------------------

	% --------------------------
	\subsection{Complexity in Krylov basis}\label{subsec:kcomplexity}
	% --------------------------
	
	\paragraph*{Time-independent Hamiltonian system.---}
	
	The idea of complexity in the Krylov basis was first introduced for operators and was later extended to states \cite{PhysRevX.9.041017,PhysRevD.106.046007}. For a detailed review, see \cite{Nandy:2024htc, Rabinovici:2025otw}. We start from an initially localized operator or state and generate an orthonormal basis by iterative action of the time evolution generator of the operators and states, namely the Liouvillean and the Hamiltonian respectively. This is best understood in case of time-independent Hamiltonian, although some instances of time-dependent Hamiltonians have been studied more recently\,\cite{PhysRevLett.134.030401}. Given a time-independent Hamiltonian $H$ and an initial state $|\psi_0\rangle$, which is also treated as the initial Krylov basis vector $|K_0\rangle$, the complexity of the final state $|\psi(t)\rangle=e^{-i H t}|\psi_0\rangle$ is defined as the average position of the state in the Krylov basis $\mathbb{K}_s := \{|K_n\rangle: n=0,1,\ldots ,\mathcal{K}_s\}$. More specifically,
	\begin{equation}
		\mathcal{C}(t)=\sum_{n=0}^{\mathcal{K}-1} n |\phi_n(t)|^2, \qquad \phi_n(t)=\langle K_n|\psi(t)\rangle.
	\end{equation}
	
	Starting from the initial vector $|K_0\rangle$, additional basis vectors are constructed using the Lanczos algorithm as follows:
	\begin{align*}
		|A_{n+1}\rangle &= (H-a_n)|K_n\rangle-b_n|K_{n-1}\rangle,\,  a_{n}=\langle K_n|H|K_n\rangle, \\
		|K_{n+1}\rangle &= \frac{|A_{n+1}\rangle}{b_{n+1}}, \quad b_{n+1}= \lVert A_{n+1}\rVert\,,
	\end{align*}
	where $\lVert \cdot \rVert$ represents the norm of the vector. While the Lanczos algorithm has a tridiagonal approximation in its construction, it works for large Hilbert spaces as well once the Full Orthogonalization is applied\,\footnote{While in case of full orthogonalization one subtracts out the overlaps with all previous basis vectors instead of just the previous two, one still focuses mostly on the two sets of coefficients $\{a_n\}$ and $\{b_n\}$.}. This recursive algorithm is stopped when the $b_n$ coefficients come to zero indicating the full Hilbert space has been explored and there is no linearly independent basis left to be explored. For state space, the Krylov space dimension $\mathcal{K}_s$ is simply the state Hilbert space dimension $d$ for ergodic dynamics, whereas for operators, it is upper bounded by $\mathcal{K}_o\leq d^2-d+1$\cite{Nandy:2024htc, Rabinovici:2025otw}. In addition, it has been shown in ref.~\cite{PhysRevD.106.046007} that the complexity is always minimum on the basis generated by such a Gram-Schmidt orthogonalization scheme. 
	
	Once we have all the basis vectors, we can proceed and compute all the $\phi_n(t)$ simply as the overlaps of the state at a given time $t$ with all the basis vectors $\mathbb{K}_s$. The significance of this particular probe, namely complexity in the Krylov basis, is that it provides us with a systematic way to measure the delocalization of the state or operator in the corresponding Hilbert space along time evolution. Hence, it has been particularly useful in the scenarios when either the evolution is chaotic, where fastest delocalization results in maximum possible complexity, or if the state undergoes some sort of localization when the delocalization slowdown results in suppressed complexity.
	
	\paragraph*{Discrete-time quantum system.---} As mentioned previously, the method of generating the basis by the action of the Hamiltonian works best if the Hamiltonian is time-independent. However, in most practical scenarios of quantum circuit constructions, one actually has control over the different unitaries being applied in the circuit at different times rather than the Hamiltonian. It was suggested in \cite{PhysRevD.106.046007} that this definition of complexity can still be applied in such scenarios if one generates a similar orthonormal basis by the action of the unitaries at different timesteps. We take their suggestion seriously and perform a detailed study on a range of random quantum circuits with discrete time dependent unitaries in this paper. Also note that it was further shown in \cite{PhysRevD.106.046007} that the following way of constructing the basis for a given discrete time-dependent unitary evolution minimizes the complexity over any other possible choice of basis. It is also worth clarifying that the unitary dynamics that we study in this paper do not necessarily arise from some underlying physical Hamiltonian and hence the notion of complexity depends on the protocol on the basis of which the circuit is constructed, e.g., type of randomness, periodicity, disorder strength etc. Our study shows how details of circuit construction results in different behaviour of the proposed notion of complexity and how it compares to the Krylov spread complexity for Hamiltonian evolution~\footnote{In circuit models, the layer structure is part of the physical definition of the dynamics. For instance, brickwork circuits, monitored circuits, and Floquet circuits are all specified at the level of discrete layers. In that context, changing the grouping of steps changes the protocol itself, just as \textit{changing the gate set changes ordinary circuit complexity}. Therefore, we do not claim invariance under arbitrary coarse-graining; rather, the complexity is defined relative to a chosen discrete protocol.
	}.

	Let's say we have a discrete-time quantum circuit where we denote the unitary at step $t$ as the one being applied on the state $|\psi_{t-1}\rangle$ after step $(t-1)$ by $U_t=U(t;t-1)$. Starting from the initial state $|\psi_0\rangle=|K_0\rangle$, we generate the orthonormal basis $\mathbb{K}_s$ in the following way.
	\begin{enumerate}
		\item First act on the initial state with the first unitary $U_1$ and generate the second Krylov vector as follows, 
		\begin{align*}
			|\psi_1\rangle & = U_1 |\psi_0\rangle, \\
			|A_1\rangle &= |\psi_1\rangle-\langle \psi_0|\psi_1\rangle |\psi_0\rangle,\, \\
			|K_1\rangle &= \frac{|A_1\rangle}{\lVert A_1\rVert}.
		\end{align*}
		Note that $\phi_{n,0}=\langle K_{n}|\psi_0\rangle=\delta_{n,0}$. Hence, initially the complexity starts from $\mathcal{C}(t=0)=0\times |\phi_{0,0}|^2=0$. Similarly at time $t=1$, there are simply two basis vectors $|K_{0}\rangle$ and $ |K_1\rangle$, therefore, only two overlaps $\phi_{0,1}$ and $\phi_{1,1}$ are nonzero, making the complexity $\mathcal{C}(t=1)=|\phi_{1,1}|^2$.
		\item Generalization of the first step to $n$-th step is fairly simple and looks like the following,
		\begin{align}
			|\psi_n\rangle &= U_n|\psi_{n-1}\rangle, \\
			|A_n\rangle &= |\psi_n\rangle-\sum_{i=0}^{n-1} \langle K_i|\psi_n\rangle |K_i\rangle,\\  |K_n\rangle &=\frac{|A_n\rangle}{\lVert A_n\rVert}\,. 
		\end{align}
		Again, there exists $(n+1)$ orthonormal basis vectors upto this step and hence $(n+1)$ nonzero overlaps $\{\phi_{i,n}\}_{i=0}^n$ contributing non-trivially to the complexity at timestep $n$. Note that, since the generator at each step is different in this case ($U_n$), unlike the Hamiltonian evolution, we do not have a single tri-diagonal representation of the overall unitary in this case. Furthermore, given a state $|\psi_n\rangle$ at a certain timestep $n$, we subtract the overlaps with all previous Krylov basis vectors $|K_i\rangle$ with $0\leq i\leq n-1$ to form the Krylov vector $|K_n\rangle$. Hence, unlike the Hamiltonian evolution case, the $\phi_{i,n}=\langle K_i|\psi_n\rangle$ have information of overlap of the state with all previously successive Krylov vectors $|K_i\rangle$. The complexity can be written as 
		\begin{equation}
			\mathcal{C}(t)=\sum_{n=0}^{\mathcal{K}_s-1} n|\phi_{n,t}|^2\,.
		\end{equation}
	\end{enumerate}
	The Krylov space dimension $\mathcal{K}_s$ is equals to the Hilbert space dimension $d$ for ergodic dynamics. Since the unitary evolution operator $U_t$ itself is used to generator of the Krylov basis, the complexity can grow at most linearly with time\footnote{This is previously shown in\,\cite{PhysRevB.111.014309,Sahu:2024kho} for the similar notions of complexity in which krylov basis is generated by the unitay operator.}. It follows that since the expansion coefficients $\phi_{n,t}=0\ \forall n>t$. Furthermore, the optimal growth complexity is attained for the choice $\phi_{n,t} = \delta_{n,t}$, which corresponds to the case where at each time-step the state evolves in an orthonormal direction.

	This is the main methodology that we implement in all the calculations done in the paper. It is worth noting that while we generate the orthonormal basis by the action of the unitaries instead of the Hamiltonian, we still call it the Krylov basis as per the suggestions made previously in \cite{PhysRevD.106.046007}. In appendix \ref{appendix:relation-complexity}, we discuss the relation between the Hamiltonian and unitary based complexities. In the next section, we describe the models in which we study this notion of complexity for state evolution under various discrete-time-dependent setups.
	
	% -------------------------
	\subsection{Models}\label{subsec:models}
	% -------------------------

	We consider discrete time evolution of two classes of one-dimensional quantum circuit models --- Random unitary circuits (RUCs) and Floquet unitary circuits (FUCs).

	\subsubsection{RUCs}\label{subsec:model_ruc}
	
	\begin{figure}
		\includegraphics[width=0.85\linewidth]{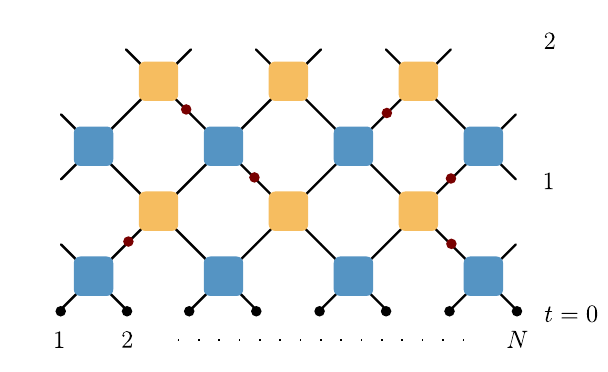}
		\caption{Illustration of RUC model consists of $N$ qubits. The bricks represent local unitary operation arranged in brickwork manner. A single evolution steps consists of two-layers \textit{i.e.} odd (blue) and even (yellow) of unitaries. The red circles represents the local measurement on qubit, while the one in blue represents initial reference state.} 
		\label{fig:C_RUC}
	\end{figure}
	
	In RUCs, for each discrete time step $t =1,2,\ldots,$ there is a unitary evolution operator $U_t =U(t;t-1)$ under which a pure state evolves as $|\psi_t\rangle = U_t |\psi_{t-1}\rangle$. We take $U_t$ to be a product of local two-qubit gates arranged in a bricklayer pattern (see Fig.~\ref{fig:C_RUC}, left). Explicitly,
	\begin{equation}
		\begin{split}
			U_t &= \left(\prod_{x\in \text{odd}} \mathcal{U}_{x,2t+1} \right)\left(\prod_{x\in \text{even}} \mathcal{U}_{x,2t}\right) \equiv \mathcal{U}^{(o)}_t \cdot  \mathcal{U}^{(e)}_t
		\end{split}
	\end{equation}
	where $\mathcal{U}_{x,\tau}$ is the gate acting on qubit pair $x$ at time step $\tau$. Each local unitary gates $\mathcal{U}_{x,t}$ is an independent \textit{Haar-random} unitary drawn from $\mathsf{U}(4)$. Even the minimal brickwork circuit described above retains essential feature of spatial locality of the interaction, which is natural in quantum information and relevant for toy models of black holes\,\cite{patrick_hayden_black_2007,yasuhiro_sekino_fast_2008,lashkari_towards_2013,shenker_stringy_2015}. The quantum circuit complexity of such system grows linearly with time before saturating. The saturation occurs when the number of applied gates reaches a threshold that scales exponentially with the number of qubits\,\cite{haferkamp_linear_2022}. Moreover, their scrambling dynamics resemble those in large-$N$ or semiclassical systems, but the growth of the OTOC exhibits a diffusively broadened wavefront i.e. the crossover region of the ballistically expanding operator growth\,\cite{PRXQuantum.5.010201}.

	In addition to purely unitary evolution, it is common in studies of RUCs to incorporate nonunitary dynamics by introducing local projective or generalized measurements at random spacetime locations. Such hybrid models generate new sets of questions and give rise to phenomena with no traditional analog, such as emergent dynamical phases in quantum systems that are continuously monitored by an external observer. We implement this by ``puncturing'' the brickwork circuit with randomized single-qubit measurements, performed on a fraction $p$ of all sites. Upon a measurement, the wave function transforms as
	\begin{equation}\label{eq:measure}
		|\psi\rangle \rightarrow \frac{M_\alpha |\psi\rangle }{\lVert M_\alpha|\psi\rangle \rVert}
	\end{equation}
	where $\{M_\alpha\}$  are a set of linear \textit{generalized measurement} operators satisfying $\sum_\alpha M^\dagger_\alpha M_\alpha = \mathbb{1}$. The measurement process in Eq.~\eqref{eq:measure} is probabilistic, with outcome $\alpha$ occurring with probability $p_\alpha = \langle \psi |M^\dagger_\alpha M_\alpha |\psi\rangle$. 
	
	In our study, we choose the measurement operators to be mutually orthogonal projectors, $M_\alpha \rightarrow P_\alpha$, with $P_\pm = (1\pm Z)/2$ measure the $Z$-component of a single qubit's spin. These projectors satisfy $P_\alpha P_\beta = \delta_{\alpha \beta} P_\alpha$ and $\sum_\alpha P_\alpha = 1$. The resulting hybrid circuit undergoes a measurement-induced phase transition at a finite measurement rate $p= p_c$, separating a volume-law entanglement phase from an area-law entangled phase\,\cite{PhysRevX.9.031009}.

	\subsubsection{FUCs}\label{subsec:model_FUCs}

	In FUCs, the evolution is made time-periodic such that $U_t = U_{t+1}$, i.e., the two layers are repeated identically in the spirit of Floquet evolution. We consider several variations of Floquet evolution with a unitary circuit, which was previously shown to exhibit localized phases\,\cite{PhysRevB.98.134204}. The scenarios we consider are the following: \\
	
	\noindent (A) \textit{Gaussian circuits}: We consider $N$ fermionic pairs $(a_i,a_i^\dagger)$ satisfying $\{a_i,a^\dagger_j\} = \delta_{ij}$. We begin by defining a set of Hermitian fermionic operators given by
	\begin{equation}
		q_i = \frac{1}{\sqrt{2}} (a^\dagger_i + a_i ) \qquad \text{and} \qquad p_i = \frac{i}{\sqrt{2}} (a^\dagger_i - a_i )
	\end{equation}
	referred to as Majorana modes. Turning to the covariance matrix $\Omega$, if we choose the Majorana basis $\xi^a \equiv (q_1,p_1,q_2,p_2\ldots q_N,p_N)$, 
	\begin{equation}
		\Omega^{ab} = -i \text{Tr}(\rho [\xi^a,\xi^b])\,.
	\end{equation}
	The covariance matrix\,\cite{RevModPhys.84.621} provides a straightforward framework for discussing the corresponding group of unitary transformations for fermionic Gaussian states. In this context, we focus on Gaussian circuits that map Gaussian states to Gaussian states. The most general unitary operation acting on the covariance matrix is represented by an orthogonal transformation $O\in \mathsf{O}(2N)$. We specifically consider a subclass of these unitary transformations, generated by quadratic Hamiltonians, which correspond to special orthogonal transformations $O \in \mathsf{SO}(2N)$\,\cite{jozsa_matchgates_2008,10.5555/2011637.2011640}.
	
	With this, the time evolution operator is given by an orthogonal transformation $O\in \mathsf{SO}(2N)$ built of random two-site operations $P_i, Q_i \in \mathsf{SO}(4)$,
	\begin{equation}
		O = G\left(\bigoplus_{i=1}^{N/2} Q_i\right) G^T \left(\bigoplus_{i=1}^{N/2} P_i\right)
	\end{equation}
	where 
	\begin{equation}
		G  = \begin{pmatrix}
			0 &   & & \mathbb{1}_2 \\
			\mathbb{1}_2 & 0 & & \\
			0 & \ddots & \ddots & \\
			&   & \mathbb{1}_2 & 0 
		\end{pmatrix}
	\end{equation}
	takes care of circularly shifting $\bigoplus Q_i$ by one site, and ensures periodic boundary condition. Given the block diagonal form of the evolution operator, the two-site operations $P_i$ are coupling between sites $2i-1$ and $2i$. The time average of covariance matrix can be used to assess long-time behavior of a typical state. In ref.\,\cite{PhysRevB.98.134204}, the authors showed that the inhomogeneous evolution $(P_i \not = P_j$ and $Q_i \not = Q_j)$ exhibits Anderson localization; an initially localized impurity stays localized. On the otherhand, the homogeneous evolution ($P_i = P_j$ and $Q_i = Q_j$) results in thermalization. \\
	
	\noindent (B) \textit{Spins}: We consider a system of $N$ spin-1/2 particles evolving under the unitary
	\begin{equation}\label{eq:floquet-unitary}
		U = \left(\prod_{x\in\text{odd}} \mathcal{U}_{x}\right) \left(\prod_{x\in\text{even}} \mathcal{V}_{x}\right)
	\end{equation}
	where local unitary gates $\mathcal{U}_{x}$ are drawn from probability distributions that share a single-site Haar invariance property. In other words, for any single-qubit operators $w_i$, the transformation
	\begin{equation}
		\mathcal{U}_{x} \leftrightarrow (w_1\otimes w_2) \cdot \mathcal{U}_{x} \cdot  (w_3\otimes w_4)
	\end{equation}
	(and similarly for $\mathcal{V}_x$) leaves ensemble averages $\langle \cdots \rangle$ unchanged.

	As a first example, we take $\mathcal{U}_x$ and $\mathcal{V}_x$ to be drawn independently from the Haar measure on $\mathsf{U}(4)$. Such Haar-random two-qubit gates are maximally scrambling in the sense that they efficiently spread information initially localized on a single site across the chain. Consequntly, the system thermalizes to a locally infinite-temperature state.  For our second example, we note that any unitary in $\mathsf{U}(4)$ can be written as 
	\begin{equation}\label{eq:MBL-unitary}
		(w_1\otimes w_2) \cdot \exp\left(ia\sigma_x\otimes \sigma_x + ib \sigma_y\otimes \sigma_y + ic \sigma_z \otimes \sigma_z\right)\cdot (w_3\otimes w_4)
	\end{equation}
	where $w_i \in \mathsf{U}(2)$ and $a,b,c\in \mathbb{R}$. The ensemble is defined by drawing each $w_i$ independently from the Haar measure on $\mathsf{U}(2)$ and sampling $a,b,c$ uniformly from the interval $[-h,h]$. When the interaction strength $a,b,c$ are drawn from a narrow range --- corresponding to small $h$ --- the resulting dynamics are weakly entangling and predominantly diagonal in the computation basis, effectively mimicking the action of a strong random potential. Such conditions are known to favor MBL, which typically requires disorder strong enough to overcome coupling-induced thermalization\,\cite{annurev:/content/journals/10.1146/annurev-conmatphys-031214-014726,RevModPhys.91.021001}. As shown in \cite{PhysRevB.98.134204}, the model undergoes MBL transition at the critical value $h_c \approx 0.3$ below which it exhibit MBL phase.
	% ------------------
	\section{Results}
	% -------------------

	In this section, we present our results for the complexity in Krylov basis for the various quantum circuit models introduced in Sec.\,\ref{subsec:models}.
	
	\subsection{RUCs}\label{subsec:results-RUCs}
	
	We start with considering a more generic class of RUCs in which, at each time step $t$, the evolution operator $U_t$ is an independent Haar-random unitary drawn from $\mathsf{U}(d)$, where $d = 2^N$ is Hilbert space dimension of an $N$-qubit system. Starting from an initial state $|\psi\rangle$, the dynamics generate a sequence of states 
	\begin{equation}\label{eq:evolved-states}
		\{|\psi_0\rangle, |\psi_1\rangle, |\psi_2\rangle,\ldots \} =\left\{|\psi\rangle ,U_1|\psi\rangle,U_2U_1|\psi\rangle,\ldots  \right\}\,. 
	\end{equation}
	For the two states $|\psi_n\rangle$ and $|\psi_m\rangle$ at different times such that
	\begin{equation}
		|\psi_n\rangle = U_nU_{n-1}\cdots U_1|\psi_0\rangle
	\end{equation}
	and assume $m>n$. Then
	\begin{equation}
		|\psi_m\rangle = V|\psi_n\rangle, \qquad V:=U_m U_{m-1}\cdots U_{n+1}
	\end{equation}
	so the overlap is $\langle \psi_n|\psi_m\rangle = \langle \psi_n |V|\psi_n\rangle$. Thus the  overlap depends only on the incremental circuit $V$ of length $s:=m-n$ acting on the state $|\psi_n\rangle$. Since each $U_t$ is independently Haar-distributed over the full unitary group, the state $|\psi_t\rangle$ at any time step is itself a Haar-random vector in Hilbert space. Therefore, the expected magnitude of their overlap is  $\lvert \langle \psi_n |\psi_m\rangle\rvert \sim 1/\sqrt{d}$, which becomes vanishingly small for large $d$. This orthogonality reflects the fact that the dynamics rapidly explore the entire Hilbert space, producing states that are essentially uncorrelated at different time steps.\\

	\noindent Consider the set of orthonormal vectors $\{|\phi_i\rangle\}$ constructed from the evolved states in Eq.~\eqref{eq:evolved-states} calculated using recursive orthogonalization. Then\footnote{Throughout this work, $\mathbb{E}\left(\cdot \right)$ represents ensemble average over the unitaries for the given system.},
	\begin{equation}
		\mathbb{E}\left(\lvert \langle \phi_i |\psi_n\rangle\rvert^2\right) = 
		\begin{cases}
			1/d & i< n \\
			1- (n-1)/d & i = n
		\end{cases}
	\end{equation}
	where we used the normalization condition $\sum_i^n \mathbb{E}\left(\lvert \langle \phi_i |\psi_n\rangle\rvert^2\right) = 1$. It follows that
	\begin{equation}\label{eq:complexity-ruc}
		\mathbb{E}\left(\mathcal{C}(t)\right) =\sum_{i = 0}^t i \mathbb{E}\left(\lvert \langle \phi_i |\psi_n\rangle\rvert^2\right)  =  t - \frac{t(t-1)}{2d}, \qquad t < d
	\end{equation}
	It follows that for $d\rightarrow \infty $, the complexity grows linearly with time. In finite system, the complexity saturates at exponentially late times $t_\text{sat} = d-1$ to a value $\mathbb{E}\left(\mathcal{C}(t_s)\right) = d/2$, which is generic feature of circuit complexity in chaotic systems\,\cite{https://doi.org/10.1002/prop.201500092,caputaSpreadComplexityTopological2023a}. In ref.\,\cite{caputaSpreadComplexityTopological2023a} (also see ref.\,\cite{Camargo_2024}), the authors suggested for the system that evolves under chaotic Hamiltonian, the peak overshooting, followed by a downward slope (apart from the exponentially large saturation time and saturation value) is also an universal characteristic of the complexity dynamics. However, this feature is absent in our complexity profile. The peak overshooting is due to spectral correlation in the eigenspectrum, however, since the RUC is constructed out of most extreme versions of time-dependent unitaries, these spectral correlation arising usually from Hamiltonian spectra are washed out (see appendix \ref{appendix:sff} for detail). As we will see in Sec.\,\ref{subsec:results_FUCs}, we recover these features in the chaotic FUCs.
	
	\begin{figure}
		\centering
		\includegraphics[width=0.8\linewidth]{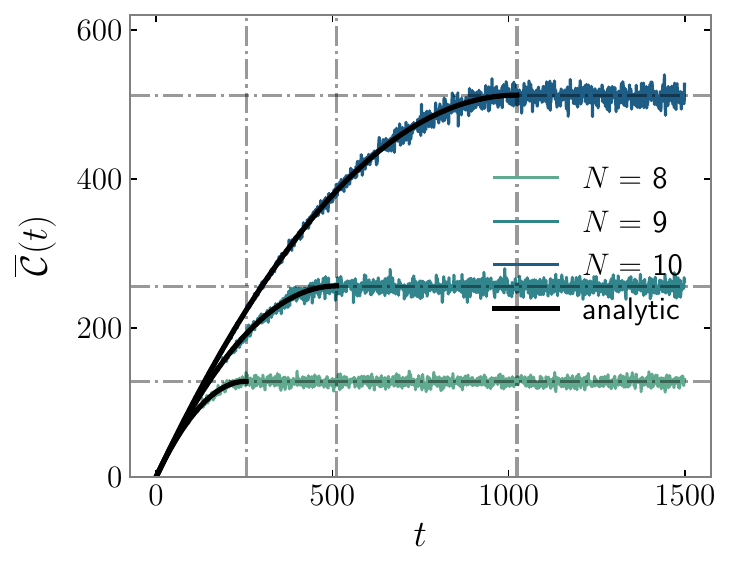}
		\includegraphics[width=0.75\linewidth]{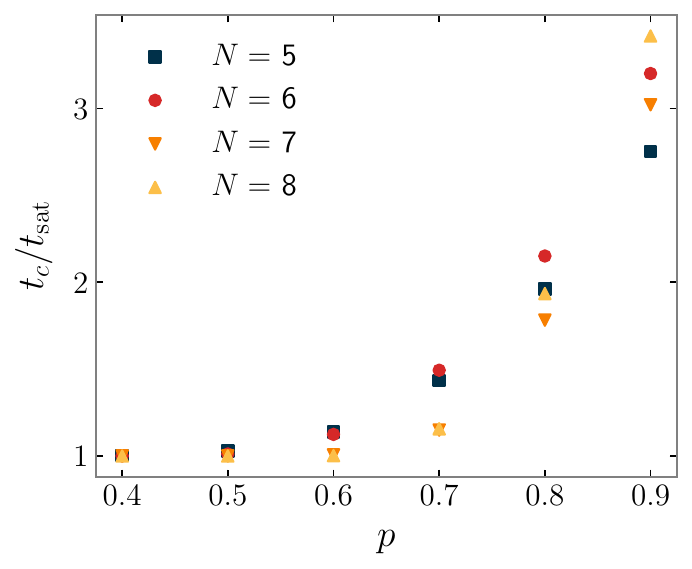}
		\caption{\textbf{Top:} The disordered complexity $\overline{\mathcal{C}}(t)$ calculated for brickwork RUCs where the local unitaries are random Haar as a function of time steps $t$. There is early time transition from linear to sublinear complexity which eventually saturates at late time. The saturation time $t_\text{sat}$ scales exponentially with system-size $t_\text{sat}\sim 2^N$ to a saturation value $\overline{\mathcal{C}}_\infty$ also exponential in system-size $\sim 2^{N}/2$. The analytical behavior in Eq.\,\ref{eq:complexity-ruc} is shown.   \textbf{Bottom}: The ratio of the time $t_c$ (defined in Sec.\,\ref{subsec:results-RUCs}) at measurement rate $p$ to that of at $p=0$, plotted as a function of $p$ computed for system-sizes $N = 5,6,7,8$. The average is taken over $2000$ copies.}
		\label{fig:comp-ruc}
	\end{figure}
	
	\vspace{5mm}
	
	\noindent \textit{Numerics.---} 
	We will now consider the numerical results obtained for RUCs with bricklayer structure introduced in Sec.~\ref{subsec:model_ruc}. In Fig.~\ref{fig:comp-ruc} (top), we show numerically computed complexity in Krylov basis for the system-sizes $N=8, 9, 10$ for RUC with bricklayer structure. The local two qubits $\mathcal{U}_{x,\tau}$ gates are independent Haar-random unitary drawn from $\mathsf{U}(4)$. Unless otherwise mentioned, we will perform the average over the large number of sample copies so that quantities are well-converged. In numerics, the initial state is a product state $|\psi_0\rangle = \lvert \uparrow\rangle^{\otimes N}$. However, taking any other states produces the same feature in the complexity profile, suggesting that it is independent of the initial state. 
	
	In order to understand the complexity behavior, we recall that the overlap $\langle\psi_n|\psi_m\rangle$ depends only on the incremental circuit $V$ of length $s = m-n$.  As $s$ grows, $V$ scrambles larger and larger portions of the system and (statistically) behaves more Haar-like. Local gates only affect qubits inside a casual light cone growing linearly with $s$. Let $l(s)$ be the number of qubits significantly affect by $V$ on the state $|\psi_n\rangle$, then $l(s)\sim \min (n,2s)$. If only $l$ qubits are effectively randomized, the typical overlap scale is set by the local Hilbert-space dimension $d_l = 2^l$, therefore, the overlap decays roughtly exponentially in $l(s)$ or $|\langle \psi_n|\psi_m\rangle|\sim 2^{-l/2}$. Once $s$ is large enough that the light cone covers the whole system (in 1D this is $s\sim \mathcal{O}(n))$, the incremental circuit $V$ acts effectively on the entire system and behaves like an approximate 2-design. Then we recover our result for the Haar-random circuit over complete Hilbert space. Since we are numerically limited to small system size, our results suggest similar growth to the complete Haar-random circuit. \\

	\noindent In monitored RUCs, the Krylov basis constructed from orthogonalization of span of vectors 
	\begin{equation}
		\{|\psi_0\rangle, M\mathcal{U}^{(o)}_1 M \mathcal{U}^{(e)}_1 |\psi_0\rangle,\ldots \}
	\end{equation}
	where $M$ represents measurement operation performed at probability rate $p$ described in Eq.\,\eqref{eq:measure}. We study the time when the system explores the complete Hilbert space. In other words, the time (denoted by $t_c$) at which the Krylov basis span the complete Hilbert space. In Fig.~\ref{fig:comp-ruc} (bottom), we compute the time $t_c$ with increasing measurement rate $p$ for system-sizes $N = 5, 6, 7, 8$. We find that the system takes more and more time to acquire the complete orthonormal basis with increasing measurement rate $p$. A limiting case corresponding to $p = 1$ is studied in appendix \ref{appendix:bound} to obtain the bound for the time $t_c$. It is easy to see from Fig.~\ref{fig:comp-ruc} (bottom) that upto certain value of $p=p_\text{th}$, the saturation value remains invariant (equals to $2^N$) of the rate of measurement. This essentially means that the $y$-axis of the plot, namely $t_c/t_\text{sat}$ is 1 upto $p_{\text{th}}$. Our numerics suggest that this $p_{\text{th}}$ is close to but more precisely $\geq 0.5$ in a system-size dependent way.  The explanation for this is that after a certain rate of measurement, the system evolution is hindered by a higher rate of measurement. While this can be a version of \textit{quantum Zeno effect} resulting in a slower state evolution, the effects becoming significant only after a certain measurement rate $p_{\text{th}}$ is suggestive of a qualitative similarity to that of \cite{Suzuki_2025, PhysRevX.9.031009}.

	\subsection{FUCs}\label{subsec:results_FUCs}
	% -------------------------------------------------
	
	We now move to two classes of FUCs introduced in \ref{subsec:model_FUCs}, namely Gaussian circuits and spins. \\
	
	\begin{figure}
		\centering
		\includegraphics[width=0.49\linewidth]{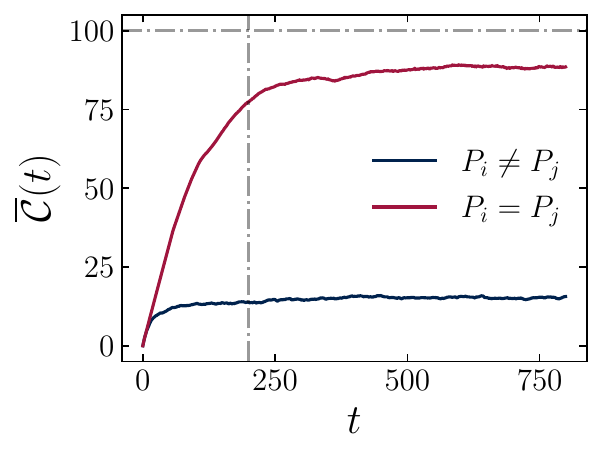}
		\includegraphics[width=0.49\linewidth]{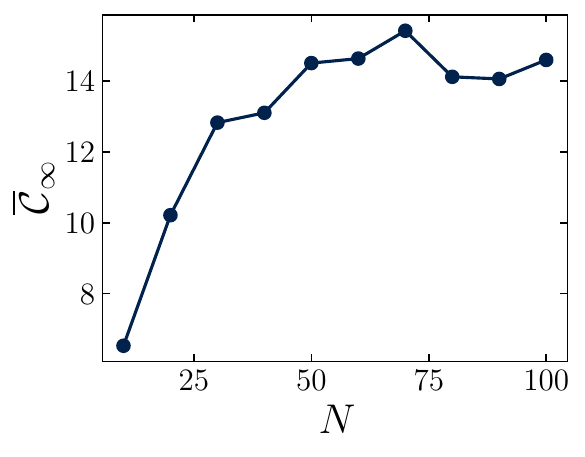}
		\caption{\textbf{Left}: The disorder-averaged complexity in Krylov basis $\overline{\mathcal{C}}(t)$ calculated for the initial product state and $N =100$ fermionic pairs. The two-site operations $P_i,Q_i$ are drawn from the special orthogonal transformation $\mathsf{SO}(4)$. The late-time saturation value is suppressed in inhomogeneous setting as result of Anderson localization. \textbf{Right}: The late-time saturation value $\overline{\mathcal{C}}_\infty$ plotted for varying system-size in inhomogeneous setting showing that the its convergence in thermodynamic limit. }
		\label{fig:complexity-gaussian}
	\end{figure}

	\noindent (A) \textit{Gaussian circuits.}--- In Fig.\,\ref{fig:complexity-gaussian}, we show the complexity in Krylov basis associated with inhomogeneous and homogeneous settings. We find that the late-time saturation value is suppressed in the inhomogeneous case due to presence of Anderson localization. The saturation value $\overline{\mathcal{C}}_\infty$ in inhomogeneous case goes to a constant in thermodynamic limit $N\rightarrow \infty$. On the other hand, although the saturation value is comparably large, it does not attains its maximum value like RUC models. Furthemore, we have also studied two other probes, namely spectral form factor and mean level spacing ratio to investigate chaotic features in this system (See sections \ref{appendix:sff} and \ref{appendix:mean-lvl-spacing}). Both the probes suggest that neither of these two models are chaotic. Hence in this particular model, as reported previously, we have a thermalization (without chaos) to localization transition distinguished by the saturation value of the complexity. This can be explained using the fact that ``Gaussian" here means quadratic in fermions (or equivalently, free models after Jordan–Wigner). These are non-interacting, and the dynamics is exactly solvable because the time evolution reduces to an orthogonal/symplectic transformation on single-particle modes. Such a spectrum factorizes into slater determinant of independent single-particle eigenmodes, so there is no many-body level repulsion (as further supported by numerics in Appendix D). To the best of our understanding, this lack of chaos is absence of chaos is directly tied to the lack of interactions and the integrable/free-fermion structure in the Gaussian model. \\

\begin{figure}
		\centering
		\includegraphics[width = 0.49\linewidth]{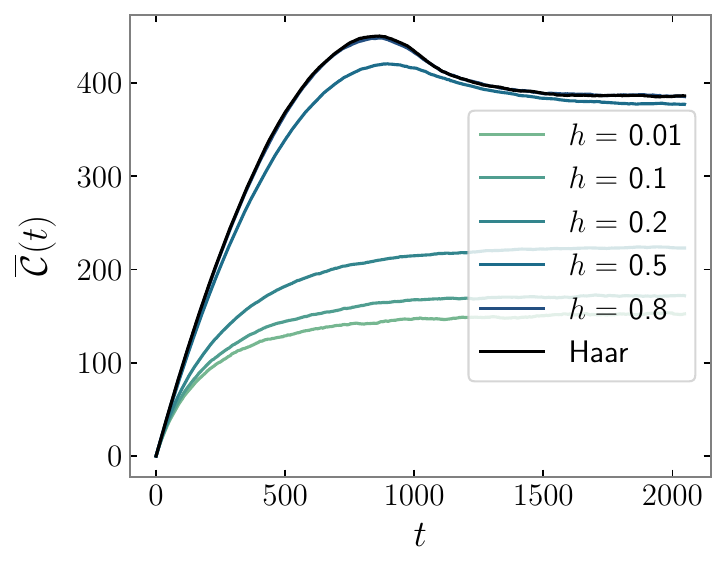}
		\includegraphics[width = 0.49\linewidth]{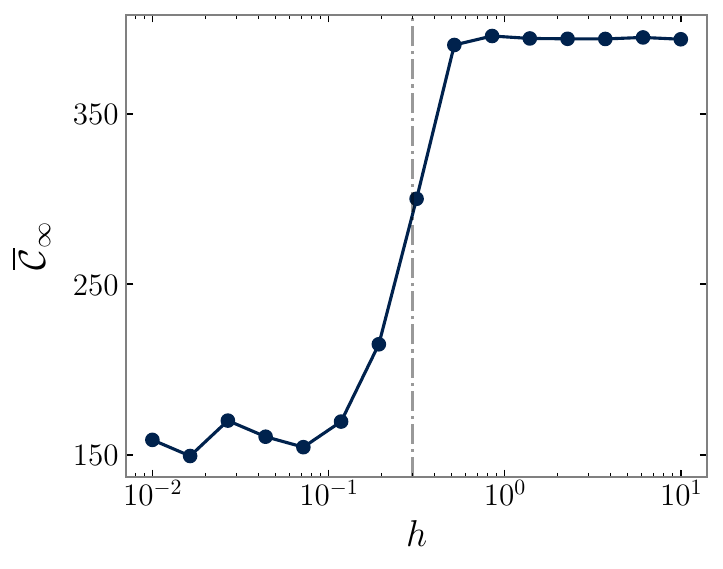}
		\caption{The disordered complexity in Krylov basis $\overline{\mathcal{C}}(t)$ for unitaries distributed according to Eq.\,\eqref{eq:MBL-unitary} with random coupling strength $h$. For strong coupling, the late-time saturation is exponential in degrees of freedom showing the thermalization of system. In contrast, the late-time saturation is suppressed at weak coupling and the system displays localization.}
		\label{fig:complexity-spins}
	\end{figure}

	\noindent (B) \textit{Spins.}--- To begin with we consider the local unitary gates $\mathcal{U}_x$ and $\mathcal{V}_x$ to be drawn from random Haar. In Fig.\,\ref{fig:complexity-spins} left, we present numerical results for complexity. The complexity profile remains similar to a random unitary case, except it features a peak before saturating.
	
	Next, we consider the tunable coupling case where the local unitaries $\mathcal{U}_x,\mathcal{V}_x$ are drawn according to the distribution in Eq.\,\eqref{eq:MBL-unitary}. In Fig.\,\ref{fig:complexity-spins} left, we show the complexity $\overline{\mathcal{C}}(t)$ for varying coupling strengths $h$. With decreasing coupling strength $h$, the late-time saturation value $\overline{\mathcal{C}}_\infty$ decreases and the downward complexity slope starts to vanish. The suppressed saturation value and absence of downward slope are signature of the MBL phase transition. To find the exact transition value, we further show the late-time saturation value with varying coupling strength in Fig.~\ref{fig:complexity-spins} right. The saturation value undergoes a transition at $h_c\approx 0.3$ in agreement with previously reported results\,\cite{PhysRevB.98.134204}. Again, we did a supporting study of SFF and MLSR of the averaged spectrum of the floquet unitary for the spin system and indeed find that in this case, the larger disorder $h\geq 0.3$ corresponds to chaotic features such as the dip-ramp-plateau of the SFF and GUE value of MLSR as opposed to no ramp and Poissonian MLSR for smaller disorder strength $h<0.3$. This also explains the presence of the peak in complexity in the spin floquet circuit for large disorder strength, which is well known to be another clear indication of chaos\,\cite{caputaSpreadComplexityTopological2023a} as opposed to no peak and suppressed complexity for small disorder corresponding to the Anderson localization\footnote{It is worth stressing that what we report regarding localization is that if there is any wavefunction localization (MBL or Anderson) in system dynamics, complexity in Krylov basis can probe that through the suppression of the complexity saturation value. This is different from the Krylov localization (more fluctuations) of Lanczos coefficients $b_n$ for integrable systems reported in \cite{rabinoviciKrylovLocalizationSuppression2022}.}.
	
	\vspace{0.2cm}
	\noindent \textbf{Connection to localization length:} In a $\ell$-qubit model with MBL, localization length $\xi$ is defined as the characteristic length scale over which correlations decay exponentially in space. For example, the connected part of the two-point correlator of two operators at $i$-th and $j$-th sites in eigenstates for such system is expected to show the following scaling behavior,
	\begin{equation}
		\langle \psi|\sigma_i^z \sigma_j^z|\psi\rangle-\langle \psi|\sigma_i^z|\psi\rangle\langle \psi|\sigma_j^z|\psi\rangle  \propto e^{-|i-j|/\xi}.
	\end{equation}
	where $\sigma^z_i$ the Pauli $Z$ at $i$-th site. More generally, small $\xi$ implies strong localization, and strong memory of initial state while weaker localization approaching ergodic/thermal phase is implied when $\xi$ becomes comparable to system size $\ell$. Now using the above logic in our examples, it is easy to argue that if the saturation value of complexity is $2^{(\ell_d-1)}$, then $\ell_d$ is proportional to the $\xi$, and upper bounded by the system size $\ell$ (as we see in the RUCs).
	
	% -----------------------------
	\section{Summary and outlook}
	% -------------------------------
	%
	In this paper, we study the state complexity in Krylov basis for various classes of random unitary circuits. We show that the proposed notion of complexity is sensitive to the circuit construction protocol. In Haar random unitary brickwork circuits, the complexity undergoes a transition from linear to sublinear growth, before saturating which scales exponentially with the number of qubits. Interestingly, the exponential scaling of saturation time is also found in quantum circuit complexity, indicating the possible connection between the two complexities\,\cite{PhysRevLett.132.160402,PhysRevA.109.022223}. When we introduce more structure (or equivalently, true dynamical features) to the circuit, we find the following. In the monitored case, the complexity growth rate in the Krylov basis decreases with increasing measurement rate $p$, resulting in an increased saturation time. In Floquet unitary classes, the complexity probes thermal and local phases through the late-time saturation value. More specifically, inhomogeneous Gaussian circuits exhibit Anderson localization, while strongly coupled spins show many-body localization. The complexity of the operator can be studied equally well using this method, and it is also often used to probe thermal and local phases. Similarly to state complexity, the operator complexity is expected to undergo a linear to sublinear transition, before saturating exponentially large times $t_\text{sat}\sim 4^{N}$ to a value that grows exponentially with system size $\sim 4^N/2$ (see section \ref{appendix:operator-construction} for basic construction). \\
	
	\noindent In the following, we summarize the main findings of our work. We study 
	\begin{itemize}
		\item  how complexity in Krylov basis can be defined for discrete time-dependent random circuits without the presence of an explicit Hamiltonian. In doing so, one needs to recursively construct orthonormal basis vectors and compute the overlap between the state after a certain layer with corresponding basis vectors treated as the wavefunctions in the Krylov basis. 
		\item how the complexity profile, saturation values, and saturation times scale with the system size, indicating a connection with the circuit complexity. We find that in the effectively kinematic picture with Haar random quantum gates, the complexity in Krylov basis from unitary generators has the same scaling behaviour as that of the circuit complexity (linear growth and saturation of complexity at $2^{N-1}$ in time $2^N$ for $N$ qubits) and also the saturation value of complexity ($2^{N-1}$) agrees with Krylov complexity saturation for Hamiltonian evolution ($\mathcal{K}_s/2$ if the Krylov space dimension is $\mathcal{K}_s$). Note that in case of state evolution through unitary generators, the Krylov basis is $\mathcal{K}_s/2=2^N$. %\himanshu{I changed here the notation for Kyrlov dimension consistent with Introduction.}
		
		\item how the complexity profile changes under randomized measurements. We find that with increasing measurement rate, after a certain measurement rate threshold based on system-size $p_{\text{th}}$, the complexity saturation time goes through a sharp increase. For a larger system size $N$, beyond $p_{\text{th}}$, the ratio $t_c/t_\text{sat}$ is greater.
		
		\item In Floquet setups, we show how this notion of complexity in the Krylov basis, generated by the unitaries in the discrete-time picture, acts as a probe of various kinds of localization (Anderson and MBL) through suppression of the saturation value. This notion of complexity can therefore measure the localization length of the wave function in the state Hilbert space.
		
		\item We focus on the random circuit-based models (RUC being fully randomly time-dependent and FUC being repeating one random realization periodically) where all our results are averaged over many different realizations. While we also find, similar to previous studies that the chaotic models have higher saturation values as compared to the localized ones, we further find presence of a clear peak before saturation of complexity in chaotic time-periodic floquet example which was not discussed in ref.\,\cite{Nizami1operatorfloquet,Nizami2spreadfloquet} (similar to \cite{PhysRevD.106.046007, Camargo_2024}). We have also supported our arguments using other probes (SFF and MLSR) presented in the appendices \ref{appendix:sff}, and \ref{appendix:mean-lvl-spacing}. While \cite{Nizami2spreadfloquet} also studied MLSR in certain floquet models, our SFF studies are novel in these circuit models of floquet evolution.
	\end{itemize}
	There are a number of questions that remain to be answered in future works. A pressing one is if its possible to probe scrambling transition in the monitored RUCs more precisely using complexity. The definition of complexity allows it to grow even in the absence of any growth in entanglement. For example, a discrete evolution in which a product state evolves to another product state results in the growth of complexity, but not entanglement. A possible way could be to consider radiative random unitary circuits\,\cite{PhysRevLett.131.220404} which were previously shown to exhibit a scrambling transition probed using OTOCs. It would also be interesting to explore complex circuit architectures beyond brick-wall and Floquet random circuits, such as nonlinear circuits or circuits with varying connectivity~\cite{nielsen2024programmablenonlinearquantumphotonic, holmes2023nonlineartransformationsquantumcomputation}. This can provide insights into how spatial arrangement and connectivity affect complexity growth and localization properties. One can also investigate how the introduction of long-range interactions alters the complexity dynamics in comparison to short-range interactions~\cite{Block_2022, Defenu_2023}. This can provide valuable insights into the transition behaviors and complexity saturation points. Further, one can examine the implications of our findings for quantum error correction and quantum computing to understand how complexity growth can inform the development of efficient quantum algorithms or new quantum-based error-correcting codes, particularly in the context of noisy intermediate-scale quantum (NISQ) devices and assess how complexity can be harnessed for tasks like quantum state discrimination, where the structure of Krylov basis can potentially inform more efficient strategies~\cite{NISQPhysRevResearch.3.013063, cindrak2024measurablekrylovspaceseigenenergy, cindrak2025krylovcomplexityobservabilitycapturing}. Finally, it can be worth analyzing the transport properties of qubit systems in and out of localization regimes and implications of complexity saturation on transport phenomena such as thermal conductivity or charge transport in disordered systems~\cite{Setiawan_2017}.\\

	\noindent \textit{Acknowledgement.}--- We would like to thank Vijay Balasubramanian, Sumilan Banerjee, Mario Flory, Shane Kelly, Subroto Mukherjee, and Zahra Raissi for useful discussions. AB acknowledges support from United Kingdom Research and Innovation (UKRI) under the UK government’s Horizon Europe guarantee (EP/Y00468X/1). A.B.~was previously supported by the Polish National Science Centre (NCN) grant 2021/42/E/ST2/00234 (till 31.08.2025). This research was supported in part by the International Centre for Theoretical Sciences (ICTS) for participating in the program - ``Quantum Information, Quantum Field Theory and Gravity" (code: ICTS/qftg2024/08). Research at Perimeter Institute is supported in part by the Government of Canada through the Department of Innovation, Science and Economic Development and by the Province of Ontario through the Ministry of Colleges and Universities.\\

	\noindent\textit{Code Availability.}--- The code to reproduce results of this paper are openly available in the \href{https://github.com/himanshu-PI/haar-comp}{\textit{haar-comp}} github repository and archived in Zenodo\,\cite{sahu_2025_complexity}.

	\appendix

	\section{Hamiltonian and unitary based complexity}\label{appendix:relation-complexity}
	% ==============================

	In Sec.\,\ref{subsec:kcomplexity}, we define complexity for a discrete-time quantum circuit. This complexity depends on the unitaries $\{U_n\}_n$, or equivalently on the states $\{|\psi_n\rangle\}_n$. It is natural to ask how this notion relates to Hamiltonian-based complexity. Addressing this question requires specifying an underlying Hamiltonian dynamics. In the absence of such a specification, there are arbitrary choices of  Hamiltonian. Since each unitary $U_i$ can be viewed as a curve in $\mathsf{SU}(d)$ generated by a Hamiltonian $H_i$ (possibly time-dependent) via the Schr\"odinger equation $\dot{U}_i = -i H_i U_i$, we choose, among all such possibilities, the Hamiltonian that minimizes the complexity. 
	
	More specifically, consider the decomposition of the state vector 
	\begin{equation}
		|\psi_{i + 1}\rangle = U_i |\psi_i\rangle = c_0 |K_{i + 1}\rangle + c_1 |\chi_\parallel\rangle,
	\end{equation}
	where $\chi_\parallel$ is a vector that lies within the Krylov subspace generated up to the $i$-th step. The vector $|K_{i + 1}\rangle$ is orthogonal to this subspace and thus defines the next Krylov element. In this basis, we can equivalently represent the unitary along with an associated Hamiltonian, as
	\begin{equation}\label{ref:uni-krylov}
		\widetilde{U}_i = 
		\begin{bmatrix}
			c_0 & c^*_1 \\
			c_1 & - c^*_0
		\end{bmatrix} 
		= e^{-i\delta t \widetilde{H}_i }\,.
	\end{equation}
	Here, we assume that the Hamiltonian $\widetilde{H}_i$ acts on the system for a time interval $\delta t$. The unitary (and corresponding Hamiltonian) is expressed in the Krylov basis $\left\{|K_{i + 1}\rangle, |\chi_\parallel\rangle\right\}$. We emphasize that this Hamiltonian need not be physically local, in contrast to other notions of complexity such as Nielsen complexity\,\cite{nielsenQuantumComputationGeometry2006,dowlingGeometryQuantumComputation2008}. 
	
	Now suppose that  the unitary evolution is generated by an underlying Hamiltonian. As discussed in Ref.\,\cite{PhysRevB.111.014309, Sahu:2024kho, Beetar:2025mdz}, the complexity defined via unitaries begins to deviate from the Hamiltonian-based complexity at longer times. To see this, consider the state
	$|\psi_1\rangle = U|\psi_0\rangle$ where 
	\begin{equation}
		U = e^{-i H \delta t} 
		= \sum_k^\infty \frac{(-i \delta t)^k}{k!} H^k 
		\approx  \sum_k^r \frac{(-i \delta t)^k}{k!} H^k.
	\end{equation}
	Here, $r$ is chosen such that the truncation error is sufficient small. The Hamiltonian based complexity generated by the Krylov basis  $\mathbb{K}_H$ associated with $\{H^k|\psi_0\rangle : k=0,\ldots ,r\}$, whereas the unitary-based complexity is generated by the Krylov basis $\mathbb{K}_U$ obtained through orthogonalization of $\{|\psi_0\rangle , U|\psi_0\rangle\}$. The deviation between the two notions of complexity arises because $\mathbb{K}_H$ generally contains a larger set of basis elements.
	
	For sufficiently small time $\delta t$, we have $U \approx  \mathbb{I} - i\delta t H$ up to $O(\delta t^2)$. In this regime, the two set $\mathbb{K}_H$ and $\mathbb{K}_U$ approximately coincide, leading to identical early-time complexity growth. However, as time increases, both the trucation error $O(\delta t^2)$ and instability of orthogonalization procedure under small perturbation itself becomes significant, causing the two complexities to deviate.

	% ------------------------------
	\section{Construction for the operator evolution}\label{appendix:operator-construction}
	% ---------------------------------------
	
	The complexity for the discrete time-dependent systems introduced in Sec.~\ref{subsec:kcomplexity} can easily be extended for the operators\,\cite{Sahu:2024kho, PhysRevB.111.014309}. Consider the operator evolution described by $W_t = U^\dagger_{t}W_{t-1} U_t = \mathscr{U}_t[W_{t-1}]$, where $t =1,2,\ldots$ and $\mathscr{U}_t$ is the unitary superoperator given by $\mathscr{U}_t = U^\dagger_t\bullet U_{t}$. We define the Krylov basis by recursively orthogonalizing each $W_t$, starting with the seed operator $|\mathcal{W}_0)=W_0$. More explicitly,
	\begin{align}
		|A_n) &= |W_n) - \sum_{i=0}^{n-1}(\mathcal{W}_i|W_n)|\mathcal{W}_n),\\
		|\mathcal{W}_n) &= \frac{|A_n)}{\lVert A_n\rVert_o}\,,
	\end{align}
	where
	\begin{equation}
		(\mathcal{A}|\mathcal{B}) = \frac{1}{d}\text{Tr}[\mathcal{A}^\dagger \mathcal{B}],\qquad \lVert\mathcal{A}\rVert_o = \sqrt{(\mathcal{A}|\mathcal{A})}\,.
	\end{equation}
	The complexity can, therefore, be defined as 
	\begin{equation}
		\mathcal{K}(t) = \sum_{n=0}^{d^2}n|(\mathcal{W}_n| W_{t})|^2\,.
	\end{equation}

	\section{Bound on saturation time}\label{appendix:bound}
	% ------------------------------------------------------------
	
	To bound the time $t_c$ when the Krylov basis span the complete Hilbert space in case of monitored RUC, we consider the limiting case in which the probability rate is exactly one \textit{i.e.} each qubit is measured after state evolution step. Since during each evolution step, the state evolves through Haar-random unitary and then undergoes measurement, this situation is equivalent to drawing $n$ basis from the basis set $|\{i\}\rangle$ with probability $p(|\{i\}\rangle) = 1/d$. Here $|\{i\}\rangle$ with $i = 0,1,\ldots ,d-1$ represents computational basis. The saturation time corresponds to when the Krylov basis form a complete basis vector over complete Hilbert space. In other words, the number of draws required $t_s$ such that each basis vector $|\{i\}\rangle$ is drawn at least once. The probability of drawing $D$ basis vector at least once in $n$ number of drawn can be written as 
	\begin{equation}\label{eq:saturation-time}
		P(n,d) = \frac{d!}{d^n} \mathcal{S}(n,d)  
	\end{equation}
	where $\mathcal{S}(n,m)$ represents the Stirling partition number which tells the number of ways to partition a set of $n$ objects into $m$ non-empty subsets. We can set the probability $P(n,d) \approx 1- \epsilon$ for $n\sim d\log (d/\epsilon)$ where $\epsilon \in \{0,1\}$ represents approximation error. Therefore, the saturation time $t_s < \mathcal{O}(d\log (d/\epsilon))$. 
	
	The minimal complexity growth can also be deduced from this limiting case. We consider the probability of drawing $m$ basis vector at least once in $n$ number of drawn given by
	\begin{equation}
		P(n,m) = {d \choose m} \frac{\mathcal{S}(n,m)\cdot m!}{d^n}\,.
	\end{equation}
	Therefore, the complexity can be written as 
	\begin{equation}
		\begin{split}
			\text{min}\  \mathbb{E}\left(\mathcal{C}(t)\right) &= \sum_{i=0}^{t} iP(t) \\
			&\cong m_\text{max} P(t,m_\text{max})
		\end{split}
	\end{equation}
	where $m_\text{max}$ corresponds to value of $m$ for which $P(n,m)$ is maximum. To find $m_\text{max}$, we examine the ratio $P(n,m+1)/P(n,m)$
	\begin{align}
		\frac{P(n,m+1)}{P(n,m)} &= \frac{d-m}{m+1}\cdot \frac{\mathcal{S}(n,m+1)}{\mathcal{S}(n,m)}
	\end{align}
	The maximum occurs when ratio transitions from $>1$ to $<1$, i.e.,  $P(n,m+1)\approx P(n,m)$. If we consider $n$ to be large, we can further approximate $$\lim_{n\rightarrow \infty}\mathcal{S}(n,m)\sim m^n/m!$$ so that
	\begin{equation}
		\begin{split}
			\frac{P(n,m+1)}{P(n,m)} &\sim \frac{d-m}{(m+1)^2}\left(1+\frac{1}{m}\right)^n \\
			&\sim \frac{dn}{m^3}
		\end{split} 
	\end{equation}
	where, in the last step, we assumed $d\gg m\gg 1$. Therefore, $m_\text{max}\sim n^{1/3}$ leading to sub-linear complexity grow $\sim t^{1/3}$.
	
	% =========================================================
	\section{Spectral Form Factor}\label{appendix:sff}
	% =========================================================
	
	In this section, we will show that the downward complexity slope after the peak seen in spin system in Fig.~\ref{fig:complexity-spins} is result from the eigenvalue correlations, just like the ramp in spectral form factor after the dip. We consider the distribution of the elements of the spectrum of the unitary Floquet evolution operator $U$ in Eq.\,\eqref{eq:floquet-unitary}
	\begin{equation}
		\text{spect}[U] = \left\{e^{i\varphi_j}: j = 1,2,\ldots,d\right\},
	\end{equation}
	where $\varphi_j$ refer to as \textit{quasienergies}. We define the \textit{spectral form factor} (SFF) as 
	\begin{align}
		\text{SFF}(t) &:=\mathbb{E}\left[\left|\text{tr}\ U^t\right|^2\right] \\
		&= \mathbb{E}\left[\sum_{i,j = 1}^d \exp\left(i(\varphi_i - \varphi_j)t\right)\right]
	\end{align}
	Here $\mathbb{E}[\cdot ]$ is an average over an ensemble of unitaries $U$.\\
	
	\begin{figure*}[t]
		\centering
		\includegraphics[width=0.3\linewidth]{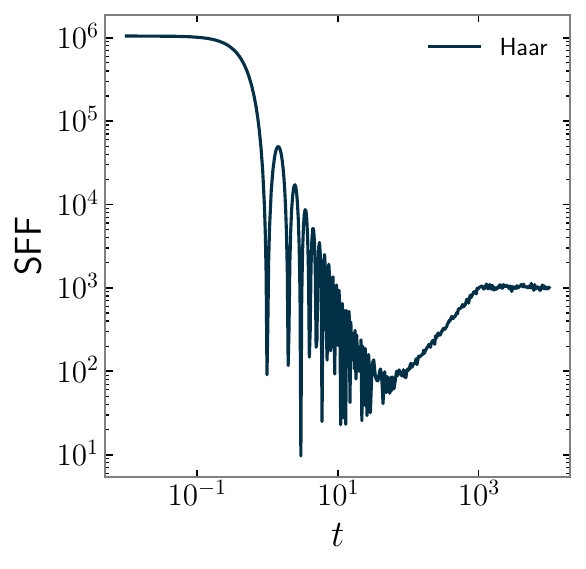}
		\includegraphics[width = 0.3\linewidth]{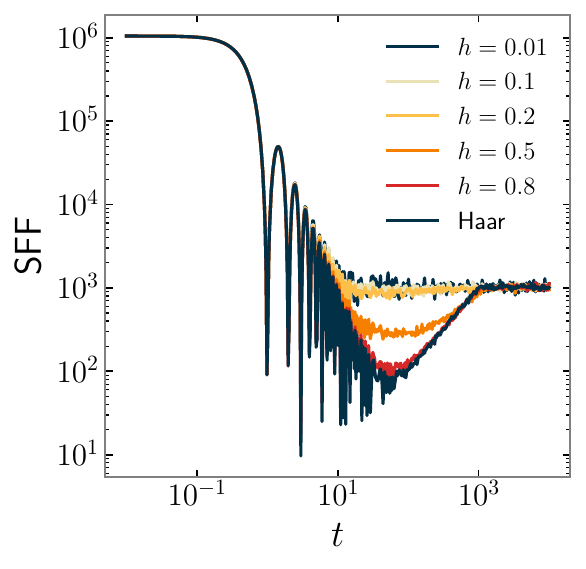}
		\includegraphics[width = 0.3\linewidth]{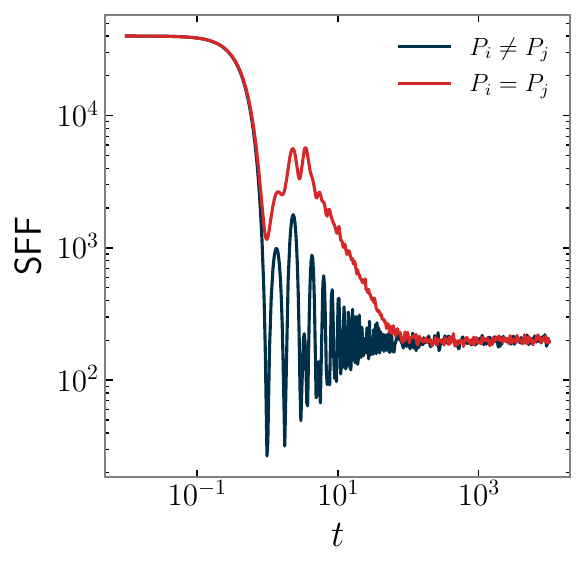}
		\caption{The SFF as a function of time $t$ in FUCs. \textbf{Left}: For the spin model, where the local unitary gates $\mathcal{U}_x$ and $\mathcal{V}_x$ are drawn independently from Haar measure on $\mathsf{U}(4)$ \textbf{Center}: For the spin model where the local unitary gates $\mathcal{U}_x$ and $\mathcal{V}_x$ take the form given in Eq.\,\eqref{eq:MBL-unitary} with  a random coupling strength $h$. \textbf{Right}: For the Gaussian circuits.}
		\label{fig:sff-floquet}
	\end{figure*}

	\noindent \textit{Spins}.--- We begin with considering the spin model where the local unitary gates $\mathcal{U}_x$ and $\mathcal{V}_x$ are drawn independently from the Haar measure on $\mathsf{U}(4)$ (see Fig.~\ref{fig:sff-floquet} (left)). The SFF follows downward slope, that lasts until the dip, followed by an upward ramp and then a plateau. The ramp is due to the repulsion between eigenvalues that are far apart in the spectrum, which is well-known in quantum chaotic systems. Therefore a linear ramp is often taken as a defining signature of quantum chaos \cite{Haake2010, PhysRevE.50.888, maldacena_bound_2016, Bertini2018,PhysRevLett.121.060601}. The spin system with local random-Haar unitary, therefore, is chaotic system. Furthermore, as argued in ref.\,\cite{PhysRevD.106.046007}, the peak and downward slope of complexity arise from the same correlation of eigenvalues similar to the ramp in SFF. This ramp from the two-level spectral correlations (equivalently SFF) match the universal predictions of RMT (the “spectral rigidity” effect). Energy levels exhibit level repulsion, eigenvalues avoid clustering, producing correlations that manifest as this ramp. Mathematically, the Fourier transform of the Wigner–Dyson two-point correlation function gives a linear growth.
	
	We can further see this connection more clearly by considering the tunable coupling model, which undergoes the MBL transition in small coupling strength $h$ limit. In Fig.~\ref{fig:sff-floquet} center, we show the SFF calculated for varying coupling strength $h$. We find that with decreasing coupling strength $h$, the ramp begins to disappear as the model moves toward the MBL phase where the level repulsion are absent. On the complexity side, this translates to the disappearance of peak and downward slope as previously observed in Fig.~\ref{fig:complexity-spins}. \\
	
	\noindent \textit{Gaussian circuits}.--- We now consider the SFF calculation for the gaussian circuits. Recall that the complexity, in this case, does not features a downward slope seen in spin model. In Fig.~\ref{fig:sff-floquet} right, we show the SFF which has similar features to that of integrable system --- downward slope followed by the plateau. We find absence of ramp in both the homogeneous and inhomogeneous case, which agree with the behavior seen in complexity profile. It is further worth noting that in the homogeneous case, while the SFF does not still show a clear dip-ramp-plateau structure, it has a strange shorter and unclear ramp and a second decay period before reaching the plateau. The unclear ramp indicates that the spectrum has some short-range correlations (slight level repulsion), but not rigid across the whole spectral window. This is characteristic of intermediate statistics, e.g. semi-Poisson (critical statistics in Anderson transitions)\cite{kravtsov1996spectralstatisticsandersontransition, Varga_2000}, or systems with partial mixing between symmetry sectors. During the second decay regime, correlations emerged previously die out again before Heisenberg time. This implies loss of spectral rigidity at long ranges in energy, typical of systems with quasi-integrability or conserved quantities that block long-range level correlations. We further calculate the mean level spacing ratio in Sec.~\ref{appendix:mean-lvl-spacing} to show that indeed both of these classes have ratio same as Poisson statistics.  \\
	
	\noindent \textit{Comment on RUCs}.---  We briefly comment on the absence of downward complexity slope in RUC model considered in Sec.~\ref{subsec:results-RUCs}. Although RUCs are minimally structured models of chaotic quantum dynamics capturing universal features such as entanglement growth and operator spreading, the lack of downward slope in complexity is consequence of the absence of spectral correlations. In contrast to spin FUC model, the intrinsic time-dependence of RUCs with independently drawn layers prevents any spectral correlation from arising.
	To see this, we can define SFF like quantity in a time-dependent setting via
	\begin{equation}
		\mathbb{E}\left[ \left| \text{tr} \left(U_tU_{t-1}\ldots U_1\right)\right|^2 \right]\,.
	\end{equation}
	where $U_\tau$ is the unitary acting on the $\tau$-th time step. For our argument, we will assume that each $U_i$ is an independent random-Haar element of $\mathsf{U}(d^2)$. Because Haar measure is the unique left- and right-invariant probability measure on $\mathsf{U}(d^2)$, the distribution of $V_t :=U_tU_{t-1}\ldots U_1$ is again Haar for every $t$. Therefore,
	\begin{equation}
		\mathbb{E}\left[\left|\text{tr}\ V_t\right|^2\right] = \mathbb{E}_\text{Haar}\left[\left|\text{tr}\  W\right|^2\right]
	\end{equation}
	where $W\in \mathsf{U}(d)$ is Haar-distributed. We can easily simplify using the properties of Haar-random matrices 
	\begin{align*}
		\mathbb{E}_\text{Haar}\left[\left|\text{tr}\  W\right|^2\right] &= \mathbb{E}_\text{Haar}\left[\sum_{a,b = 1}^d W_{aa}W^*_{bb'}\right] \\
		&= \sum_{a,b=1}^d \mathbb{E}_\text{Haar}[W_{aa'}W^*_{bb'}] \\
		&= \sum_{a,b=1}^{d} \frac{1}{d}\delta_{ab} = 1
	\end{align*}
	which is time-independent.
	
	\section{Mean Level Spacing}\label{appendix:mean-lvl-spacing}
	
	In Sec.\,\ref{appendix:sff}, we saw that the ramp is absent in the SFF profile for the Gaussian circuit. We argued that this is due to the lack of level repulsion in the eigenspectrum. To make this argument concrete, we calculate the mean level spacing ratio which is another quantity that captures this effect. Below, we give brief introduction to mean level spacing ratio (interested reader can look refs.\,\cite{10.1007/978-3-0348-8266-8_36,PhysRevLett.52.1,PhysRevX.10.021019}).
	
	\begin{figure}
		% \begin{minipage}[c]{0.4\textwidth}
			\includegraphics[width=0.95\linewidth]{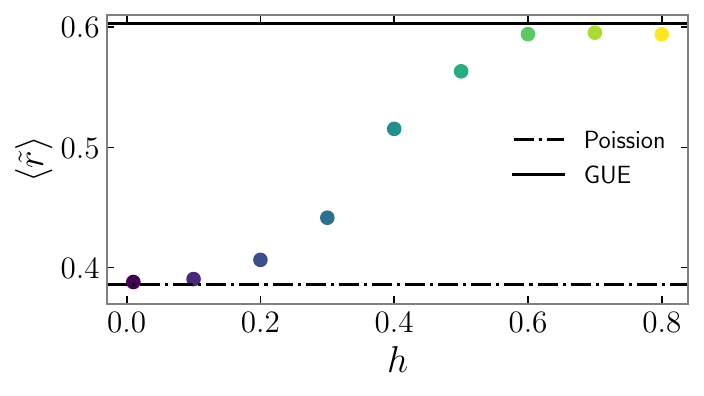}
			% \end{minipage}\hfill
		% \begin{minipage}[c]{0.5\textwidth}
			\caption{The mean level spacing ratio $\langle \tilde{r}\rangle $ calculated for the varying coupling strength $h$ calculated for the system-size $N= 10$. With increasing coupling strength $h$, the system transitions from the Poission statistics to GUE akin to the MBL transition.} \label{fig:mls-tun}
			% \end{minipage}
	\end{figure}
	
	Consider $e_n$ be an ordered set of energy levels and the nearest-neighbor spacing $s_n = e_{n+1}-e_n$. We define ratio $\tilde{r}_n$ as
	\begin{equation}
		\tilde{r}_n  = \frac{\min(s_n,s_{n-1})}{\max(s_n,s_{n-1})} = \min\left(r_n,1/r_n\right)
	\end{equation}
	where $r_n =s_n/s_{n-1}$. For quantum Hamiltonian associated with integrable classical systems, the Berry-Tabor conjecture\,\cite{10.1007/978-3-0348-8266-8_36} suggests that their level statistics follows a Poisson distribution for which $\langle \tilde{r}\rangle \approx 0.38629$. On the other hand, for quantum Hamiltonian whose classical counterparts are chaotic, the Bohigas-Giannoni-Schmit conjecture\,\cite{PhysRevLett.52.1} proposes that their level statistics should fall into one of three classical ensembles of random matrix theory (RMT) known as Wigner ensembles. For our discussion, we focus on the Gaussian Unitary Ensemble (GUE) which corresponds to systems with complex random variables as matrix entries for which $\langle \tilde{r}\rangle = 0.60266$.

	In Gaussian circuit, we find that the calculated mean level spacing ratio $\langle \tilde{r}\rangle$ to be $\approx 0.4030$ and $\approx 0.3977$ for the inhomogeneous and homogeneous case, respectively. It suggest that the our Gaussian circuit model obeys Poisson statistics, which explain the absence of ramp in the calculated SFF. On the other hand, for spin system with local Haar-random unitaries, we find $\langle \tilde{r}\rangle \approx 0.5989$ calculated for the system size $N = 10$. Therefore, the system obeys the GUE statistics, hence the ramp observed in the SFF profile. In Fig.~\ref{fig:mls-tun}, we show the $\langle \tilde{r}\rangle$ for the tunable coupling spin system. We find that with the system statistics transit from GUE to Poisson with decreasing coupling strength $h$ as result of MBL phase transition\,\cite{PhysRevB.105.174205}.

	% \bibliographystyle{apsrev4-1}
	%	\bibliography{bib}


\end{document}